\documentclass[11pt,preprint,preprintnumbers,amsmath,amssymb,nofootinbib,aps]{revtex4}
\usepackage{graphics,color,array,dcolumn}
\usepackage{hyperref}
\usepackage{calc}

\usepackage{amsmath}
\usepackage{amssymb}
\usepackage{xspace}

\usepackage{graphicx}
\usepackage{amsfonts}

\newcommand{\be}{\begin{equation}}
\newcommand{\ee}{\end{equation}}

\newcommand{\ie}{{\it i.e.}}

\newcommand{\calo}{{\cal O}}

\newcommand{\STr}{{\rm STr}}

\begin{document}

\title{Gravity from a Particle Physicists' perspective
\footnote{Lectures given at the  Fifth International School on Field Theory and Gravitation, 
Cuiab\'a, Brazil, April 20-24 2009.}
}

\author{Roberto Percacci\footnote{\it on leave from SISSA, via Beirut 4, I-34151 Trieste, Italy.
Supported in part by INFN, Sezione di Trieste, Italy}}
\email{rpercacci@perimeterinstitute.ca}
\affiliation{Perimeter Institute for Theoretical Physics, 31 Caroline St. North, Waterloo, Ontario N2J 2Y5, Canada}

\begin{abstract}
In these lectures I review the status of gravity 
from the point of view of the gauge principle and renormalization,
the main tools in the toolbox of theoretical particle physics.
In the first lecture I start from the old question ``in what sense is gravity a gauge theory?''
I will reformulate the theory of gravity in a general kinematical setting
which highlights the presence of two Goldstone boson-like fields,
and the occurrence of a gravitational Higgs phenomenon.
The fact that in General Relativity the connection is a derived quantity
appears to be a low energy consequence of this Higgs phenomenon.
From here it is simple to see how to embed the group of local frame transformations
and a Yang Mills group into a larger unifying group, and how the distinction
between these groups, and the corresponding interactions, 
derives from the VEV of an order parameter.
I will describe in some detail the fermionic sector of a realistic ``GraviGUT'' with
$SO(3,1)\times SO(10)\subset SO(3,11)$.
In the second lecture I will discuss the possibility that the renormalization 
group flow of gravity has a fixed point with a finite number of attractive directions.
This would make the theory well behaved in the ultraviolet, and predictive,
in spite of being perturbatively nonrenormalizable.
There is by now a significant amount of evidence that this may be the case.
There are thus reasons to believe that quantum field theory may eventually
prove sufficient to explain the mysteries of gravity.
\end{abstract}

\maketitle

\section{Motivations}

Our understanding of particle physics is based on two pillars: the gauge principle
and renormalization theory.
The gauge principle is the statement that at a fundamental level the interactions
between particles are mediated by vector bosons, whose dynamics is invariant
under an infinite dimensional group of local transformations.
It has a very long history, originating from Weyl's early work on a unified theory of
gravity and electromagnetism based on invariance under local scale transformations
\cite{weyl}. Weyl's theory was not viable, but through the work of Fock, London,
Pauli, Yang and others, its developments ultimately led to the formulation of 
nonabelian gauge theories and then to their successful application in the standard model. 
For a fascinating account of the history of this line of thought see e.g. \cite{history}.

The development of gauge theories is intimately connected to the search for unification,
but the standard model is itself not a truly unified theory, because its gauge group
is a direct product of the the color group $SU(3)_c$, the isospin group $SU(2)_L$ 
and the hypercharge group $U(1)_Y$.
In fact, what makes the standard model somewhat nontrivial from this point of view
is just the fact that the generator of the electromagnetic $U(1)$ is a mixture of 
one generator of $SU(2)_L$ and hypercharge.
The role of the Higgs field in the standard model is mainly to pick the specific
direction in the algebra which remains unbroken.
Truly unified theories of the strong, weak and electromagnetic interactions,
also based on the gauge principle but employing a simple group, 
go under the name of Grand Unified Theories or GUTs.
The most successful GUT is based on the group $SO(10)$, and it has the very nice
property that a single 16-dimensional Weyl fermion of $SO(10)$ contains
all 15 fundamental fermions of one family, plus one (an $SU(2)_L$ singlet)
which can be interpreted as a right-handed neutrino.
See \cite{bertolini} for some recent developments.
Unfortunately the main prediction of these theories, proton decay, has not
been observed in spite of great experimental efforts, so their status
is far more speculative than that of the standard model.
Still, they do an impressive job at explaining the otherwise
seemingly arbitrary assignments of the quantum numbers of the fermions, 
so it is hard to escape the impression that they must contain some degree of truth.

Renormalization originated not from abstract theory 
but rather from the struggle to overcome a nasty technical problem.
If one supposes that spacetime is continuum, then in any finite volume of space
there is an infinite number of degrees of freedom, and in summing their
contributions to physical processes one often finds divergent, and hence meaningless results.
Renormalization originated as a technical trick to absorb these divergences
into redefinitions of the couplings:
it relates so called ``bare'' couplings, which appear in the fundamental Lagrangian
and have no direct physical significance, to ``renormalized'' couplings,
which correspond to what one actually measures in the laboratory.

Later on, thanks largely to the work of Wilson \cite{wilson},
renormalization came to be understood in more general terms.
Imagine a system consisting of a large number of oscillators with different frequencies
$\omega_i$.
When one deals with a problem which is characterized by some energy scale $E$,
one cannot directly excite the oscillators whose energy levels $E_i=\hbar\omega_i$ 
are higher than $E$.
Nevertheless, the presence of those degrees of freedom affects low energy physics:
through vacuum polarization effects, they change the effective values of the charges.
In a functional integral, one can compute the effective charges as coefficients
in an effective action which is obtained by ``integrating out'' all the degrees
of freedom with energies larger than $E$.
Consequently, the observed (renormalized) strength of the interaction between
two particles will depend on the energy of the interacting particles.
%\footnote{This effect has been measured both for the electromagnetic and strong
%interactions.}
%

It is important to observe that although the formal definition of the effective action
as the result of a functional integration inevitably involves the regularization
of divergent quantities, the {\it difference} between two Wilsonian effective actions
associated to two energy scales $E_1$ and $E_2$ is finite, because it involves
only a finite range of momenta. Symbolically:
$$
\int_{E_2}^\Lambda-\int_{E_1}^\Lambda=\int_{E_2}^{E_1}
$$
where $\Lambda$ is some UV cutoff that one would like to send to infinity.
The beta functions, being the difference between couplings in two infinitesimally 
close Wilsonian effective actions, are therefore free of UV divergences.
One could take the attitude that since only renormalized quantities can be measured,
it is never necessary to talk about the bare action, nor about a UV regulator.
From this point of view the problem of the ultraviolet divergences takes a rather
different form.
One can compute the beta functions as described above, use them
to study the dependence of the renormalized couplings on energy, 
and check whether in the limit of infinite energy divergences appear or not.
In particular it may happen that all couplings tend to a fixed point,
in which case the theory would be well behaved in the UV.
In my second lecture I will describe the application of this philosophy to gravity,
and I will provide evidence that it is better behaved than
one would normally expect.

At this point I should make a comment on the title of these lectures.
The point of view that I shall describe here is that
four dimensional quantum field theory may be enough to construct 
a quantum theory of gravity, unified with the other interactions.
This is probably still a minority point of view in the particle physics
community, but insofar as the tools that are used here are the same that
have been successfully applied to the electroweak and strong interactions,
I feel it is justified to call this ``a particle physicist's point of view''.
I should also add that almost all that I will describe here is still in an early state of 
development and, given that no experimental input is available, quite speculative.
It is possible that some of these ideas will soon be found to fail for some reason,
but if this is not the case there is a rather vast new territory to be explored.

Finally, a historical note.
Nobody would be better entitled to talk about the gauge principle than C.N. Yang.
When I was starting my research life, I attended a seminar by Yang where he described
the correspondence between the formalism of gauge theories and the mathematical
theory of fiber bundles. This deeply affected my thinking and my research interests.
It was by following this thread that I arrived at the picture of gravity that I will
describe in my first lecture.
I am therefore very sad that he will not be able to come to Cuiab\' a.
The material of the second lecture is the application to gravity of the general ideas of Wilson.
This was first discussed by Weinberg \cite{weinberg}, who introduced the term
``asymptotically safe'' to describe a quantum field theory with this kind of UV behavior.
\footnote{The term ``nonperturbatively renormalizable'' is also sometimes used.
I find that ``asymptotically safe'' is a very appropriate terminology in a
particle physics context, because it immediately suggests that this is
a generalization of asymptotic freedom.}
However, at the time the evidence
for asymptotic safety was quite scant and the idea lay dormant for almost two decades.
A practical tool for the calculation of Wilsonian beta functions in this context 
only appeared in the early nineties \cite{wetterich} and was first
applied to gravity in a seminal paper by Reuter \cite{reuter1}.

\section{The Higgs phenomenon}

The Higgs phenomenon plays a central role in modern unified theories of
fundamental interactions, so I will begin by recalling its main aspects.
For definiteness I will discuss a gauge theory of the group $SO(N)$,
with gauge field $A_{\mu ab}=-A_{\mu ba}$, $a=1,\ldots,N$,
coupled to a Higgs field $\phi^a$ in the fundamental representation.
The action contains a kinetic term quadratic in the covariant derivative
$D_\mu\phi^a=\partial_\mu\phi^a+A_{\mu ab}\phi^b$
as well as a potential
$V=\frac{\lambda}{4}(\phi^2-\upsilon^2)^2$.
The minimum of the potential 
occurs at $\langle \phi^2\rangle = \upsilon^2\not=0$.
One can choose the unitary gauge such that the field is always aligned
along the $N$-th axis. (This conditions leaves a residual gauge freedom
consisting of local $SO(N-1)$ transformations.) 
In particular in this gauge the vacuum configuration is
$\langle\phi^a\rangle=(0,\ldots,0,\upsilon)$, and the kinetic term of the scalar
field becomes
$D_\mu\phi^a D^\mu\phi^a \mapsto \upsilon^2 A_{\mu aN} A^{\mu aN}$,
{\it i.e.} a mass term for the components of the gauge field with
Lie algebra indices $aN$, with $a=1,\ldots,N-1$.
The components with $a,b=1,\ldots,N-1$, which span the subalgebra
of $SO(N-1)$, remain massless.

There is a variant of this that we may call the Higgsless Higgs mechanism.
To motivate it, observe that the space $R^N$ carrying the fundamental representation
can be divided into orbits of the group $SO(N)$, 
{\it i.e.} subspaces formed by points that can be related by group transformations
(the mathematical term for this is that the group ``acts transitively on the orbits'').
Apart from the exceptional orbit consisting only by the origin, all other orbits are 
$N-1$-dimensional spheres.
It seems natural to use coordinates in field space adapted to this group action.
Let $\rho$ denote the radius and $\varphi^\alpha$, $\alpha=1,\ldots,N-1$, 
denote coordinates on the sphere $S^{N-1}$.
The field $\rho$ is gauge invariant and is called the Higgs field proper,
while the coordinates on the sphere are called Goldstone bosons.
Without loss of generality we can choose the coordinates so that the 
``north pole'' $(0,\ldots,0,\upsilon)$ has coordinates $\varphi^\alpha=0$.
Because the group $SO(N)$ acts transitively on the spheres, the Goldstone
bosons are pure gauge degrees of freedom: one can transform any field
configuration $\varphi^\alpha(x)$ into any other by means of a local
gauge transformation $g(x)$.

Now imagine freezing the Higgs field to its VEV, so that one remains only
with the Goldstone bosons.
This can be achieved formally by taking the limit $\lambda\to\infty$, 
keeping $\upsilon$ constant.
However, it is not necessary to think of the theory in this way.
One can just think of constructing a scalar theory where the the field 
has values in $S^{N-1}$; it is called a NonLinear Sigma Model (NLSM).
The Goldstone bosons transform nonlinearly under $SO(N)$ transformations,
so the description of a NLSM is a little more complicated.
The covariant derivative of the Goldstone bosons can be written
$D_\mu\varphi^\alpha=\partial_\mu\varphi^\alpha+A_{\mu ab}K_{ab}^\alpha(\varphi)$
where $K$ are the Killing vectors generating the action of the group on the sphere.
The action for the NLSM coupled to gauge fields is then
\begin{equation}
\label{nlsmaction}
S=\frac{\upsilon^2}{2}\int d^4x\,D_\mu\varphi^\alpha D^\mu\varphi^\beta h_{\alpha\beta}
\end{equation}
where $h_{\alpha\beta}$ is the metric on the sphere written in the chosen
coordinate system. It is a nonpolynomial function of the coordinates.
Because by definition the group acts transitively on the fields
it is possible to choose a gauge in which $\varphi^\alpha$ has any
prescribed form, in particular we can choose the unitary gauge such 
that $\langle\varphi^\alpha\rangle=0$.
Since the vectors $K_{aN}$, with $a=1,\ldots,N-1$, form an orthonormal basis at the north pole, 
$\upsilon^2 h_{\alpha\beta}(\varphi)D_\mu\varphi^\alpha D^\mu\varphi^\beta 
\mapsto \upsilon^2 A_{\mu aN} A^{\mu aN}$
so the kinetic term of the Goldstone bosons is just a mass term for
the components $A^{\mu aN}$ of the gauge field.

So now we see two things. First, a gauged NLSM is just a gauge
invariant way of writing a massive gauge theory
\footnote{this is a special case of a general procedure that goes under the name
of ``St\" uckelberg trick''.}.
Second, we see that strictly speaking only the Goldstone bosons are necessary for the Higgs mechanism;
the Higgs field $\rho$, which is gauge invariant, is a mere spectator.
The geometrical reason why one prefers to have $\rho$ is that the full multiplet $\phi^a$
transforms linearly,
and the physical reason for preferring a linearly transforming multiplet of $N$ scalars,
over the nonlinearly transforming multiplet of $N-1$ Goldstone bosons, is that
perturbatively a linear scalar theory with a quartic potential is renormalizable,
whereas the NLSM is not.

%$$
%S=\int d^{d+1}x\biggl[-{1\over 2}\partial_{\mu}\phi^a\partial^{\mu}\phi^a-
%{2\Lambda\over\sqrt{\lambda}}\sqrt{V}+
%{\Lambda^2\over\lambda}\biggr]
%$$
%
%$\Lambda\!=\!\sqrt{\lambda V}$ 
%
%$$
%S=\int d^{d+1}x\biggl[-{1\over 2}\partial_{\mu}\phi^a\partial^{\mu}\phi^a-
%\Lambda(|\phi|^2-\upsilon^2)\biggr]
%$$

In spite of this, the NLSM has many application in diverse areas of physics,
including particle physics.
For our purposes the most important ones are to theories of the
strong and weak interactions.
The NLSM with values in $SU(N)$ describes the low energy behaviour of
QCD with $N$ massless quarks. 
In particular, when $N=2$ it provides a low energy phenomenological
description of the physics of pions \cite{gl}.
In this case $\upsilon$ should be identified with $f_\pi$, the pion decay constant. 

The application to weak interaction physics is somewhat less well known,
but equally important.
In this case $\varphi^\alpha$ with $\alpha=1,2,3$ are the spherical coordinates
of the complex Higgs doublet, and $\upsilon$ should be identified with 
the Higgs VEV, $\upsilon\approx$246GeV.
If we study weak interactions at momenta $p\ll \upsilon$,
particles with masses of order $\upsilon$ are effectively decoupled.
Assuming that dimensionless couplings are not too small,
this implies that we can ignore the Higgs field and the massive gauge fields.
This approximation to the standard model has been studied in \cite{ab}.

We conclude this section with some remarks.
The first is that although it should be regarded as an effective field theory,
the NLSM is still subject to quantum corrections.
The proper way of dealing with this problem 
goes under the name of ``chiral perturbation theory''.
One puts a cutoff on momentum integrations
at the energy scale at which the theory itself is expected to break down,
namely for energies of order $4\pi\upsilon$.
This energy is of the order of the GeV for strong interactions and of the TeV
for weak interactions.
These corrections produce new terms that are not
proportional to terms in the original Lagrangian.
For example, loops calculated with the action (\ref{nlsmaction}) produce terms
with four derivatives.
The coefficient of these new terms has to be fixed by experiment.
The theory remains nonetheless
predictive, because at a given energy scale and for a given precision
only a finite number of terms are needed to describe all scattering experiments,
and there are, so to speak, more experiments than parameters.

The second remark is that in the standard model the flavor group $SU(2)$ is gauged,
so if there did not exists a complex Higgs doublet (or at least the corresponding
Goldstone bosons) the pions themselves could be ``gauged away'' and the (Higgsless)
Higgs phenomenon would occur, giving the $W$ and $Z$ a mass of the order of $10^2$ MeV.
The fact that the $W$ and $Z$ have a much larger mass, and the pions are physical,
means that three out of four degrees of freedom od the Higgs field, 
namely the Goldstone modes, surely exist.

Finally we note that for momenta $p\ll\upsilon$
we can simply set to zero the components of the gauge field that acquire mass.
This can be restated in a gauge invariant way as the condition 
\begin{equation}
\label{lowenergy}
D_\mu\varphi^\alpha=0\ .
\end{equation}
We will see that very similar conditions occur in gravity, where they
completely constrain the connection.
 
\section{The Higgs phenomenon in gravity}

I will now describe a general kinematical
framework that can be used for many different theories of gravity \cite{perbook}.
Einstein's general relativity is one of them, but we will argue that
the correct dynamics describing gravity at high energy (meaning
energies of the order of the Planck scale) is probably different
and involves independent connection degrees of freedom.
The main point of this section will be to understand that gravity 
is a gauge theory where a Higgsless Higgs phenomenon is at work,
much like in the gauged NLSM discussed in the previous section.

The necessary geometrical ingredients are a four dimensional manifold $M$
and a real vectorbundle $E$ over $M$ with fibers $\mathbf{R}^4$, in the same
isomorphism class as the tangent bundle $TM$.
\footnote{even this topological restriction could be avoided if we are
willing to admit topological defects.}
We choose local bases $\{\partial_\mu\}$ in $TM$ and $\{e_a\}$ in $E$.
Then, the dynamical variables are:
\begin{itemize}
 \item a fiber metric in $E$, $\gamma_{ab}$,
 \item a linear connection in $E$, $A_\mu{}^a{}_b$,
 \item a soldering form $\theta^a{}_\mu$.
\end{itemize}
The fiber metric has a fixed signature that for now I will assume to be $+,+,+,+$,
just for notational simplicity.
None of the main conclusions would change for other signatures.
Note that although a metric is a tensor, the condition on the eigenvalues
is nonlinear, so that the space of all metrics is not a linear space.
In fact, the space of positive definite metrics in $\mathbf{R}^4$
is the coset space $GL(4)/O(4)$, so one can view a metric as a Goldstone boson.
The linear connection is, in particle physics language,
a Yang Mills field for the group $GL(4)$. 
Yang Mills fields with noncompact groups have problems at the dynamical level:
the invariant inner product in the algebra is indefinite,
and so if one uses the standard Yang Mills action
there are degrees of freedom with wrong sign kinetic term.
One will have to reckon with these issues at some stage,
but we will see that there is much to be learned if we ignore
them for the time being.
These first two ingredients are just a gauged NLSM,
and according to the remark made in the previous section they 
could be seen as a way of writing a massive gauge field in a gauge
invariant way.
What distinguishes gravity is really the third ingredient:
The soldering form $\theta^a{}_\mu$. It is subject to the constraint
det$\theta^a{}_\mu\not=0$, which implies that
it can be viewed geometrically as an isomorphism from $TM$ to $E$
(hence the name). The constraint also makes the soldering form
an intrinsically nonlinear object.

Given a connection and a metric in $E$, and an isomorphism of $TM$ to $E$,
we can construct ``pullback'' connection and metric in $TM$.
These are given by the formulae
\begin{eqnarray}
\label{pullbackmetric}
g_{\mu\nu}&=&\theta^a{}_\mu\, \theta^b{}_\nu\, \gamma_{ab}\\
\label{pullbackconnection}
\Gamma_\lambda{}^\mu{}_\nu&=&\theta^{-1}{}_a{}^\mu A_\lambda{}^a{}_b \theta^b{}_\nu
+\theta^{-1}_a{}^\mu \partial_\lambda \theta^a{}_\nu
\end{eqnarray}
So we can view the geometrical data on spacetime as derived objects,
constructed with more basic ingredients.
We can also define the covariant derivative of the metric and the
exterior covariant derivative of the soldering form:
\begin{eqnarray}
\label{nonmetricity}
\Delta_{\lambda ab}&=&%-\nabla_\lambda\gamma_{ab}= 
-\partial_\lambda \gamma_{ab}
+A_\lambda{}^c{}_a\, \gamma_{cb}
+A_\lambda{}^c{}_b\, \gamma_{ac}\\
\label{torsion}
\Theta_\mu{}^a{}_\nu&=&
\partial_\mu \theta^a{}_\nu-\partial_\nu \theta^a{}_\mu+
A_\mu{}^a{}_b\, \theta^b{}_\nu-A_\nu{}^a{}_b\, 
\theta^b{}_\mu
\end{eqnarray}
These are called the nonmetricity and torsion, respectively.

Let us now discuss the action of gauge transformations.
These consist of local changes of frame $e'_a(x)=e_b(x)\Lambda^a{}_b(x)$
and diffeomorphisms $x'(x)$.
The former are exactly local $GL(4)$ gauge transformations,
whereas diffeomorphisms arise because in gravity the metric is
not fixed a priori.
The action of these transformations on the fields is given by
\begin{eqnarray}
\label{transfsoldering}
\theta^a{}_\mu(x)&\mapsto&{\theta^\prime}^a{}_\mu(x^\prime)= 
\Lambda^{-1 a}{}_b(x)\, \theta^b{}_\nu(x)
{\partial x^\nu\over\partial x^{\prime \mu}}
\\
\label{transfmetric}
\gamma_{ab}(x)&\mapsto&{\gamma^\prime}_{ab}(x^\prime)=
\Lambda^c{}_a(x)\, \Lambda^d{}_b(x)\, \gamma_{cd}(x)
\\
A_\mu{}^a{}_b(x)&\mapsto& A^\prime_\mu{}^a{}_b(x^\prime)=
{\partial x^\nu\over\partial x^{\prime \mu}}
\bigl(\Lambda^{-1 a}{}_c(x) A_\nu{}^c{}_d(x) \Lambda^d{}_b(x)+
\Lambda^{-1 a}{}_c(x)\partial_\nu\Lambda^c{}_b(x)\bigr)
\end{eqnarray}
In mathematical terms, these transformations form a group
called the automorphism group of $E$.
The transformations for which $x'=x$ form a normal subgroup
called the vertical automorphisms of $E$, and the quotient
of all automorphisms by vertical automorphisms is the group
of diffeomorphisms of $M$. This is as in any gauge theory.
But in a theory of gravity there is a new feature:
the soldering form can be used to construct a
map from diffeomorphisms into automorphisms.
For a given diffeomorphism $x'(x)$, the corresponding
automorphism is given by
\begin{equation}
\label{injection}
\theta^a{}_\mu(x)\frac{\partial x^\mu}{\partial x^{\prime\nu}}
\theta^{-1\nu}{}_b(x')\ .
\end{equation}
So in the presence of a soldering form, the full gauge group is the semidirect product
of local $GL(4)$ (vertical) transformations and diffeomorphisms.

If we consider the transformation (\ref{transfsoldering})
we se that it is always possible to go to a gauge where
\begin{equation}
\label{metricgauge}
\theta^a{}_\mu=\delta^a_\mu
\end{equation}
If we fix this gauge, the pullback structures 
(\ref{pullbackmetric}) and (\ref{pullbackconnection})
have the same components as the original metric and connection,
so the distinction between latin and greek indices becomes
immaterial. In fact, having fixed the soldering form,
we can actually identify $E$ with $TM$.
This is what happens in ordinary textbook formulations of
general relativity, where the metric on spacetime, and perhaps a linear
connection, is used as a dynamical variable.
For this reason we shall call (\ref{metricgauge}) the metric gauge.
Also, note that it does leave a residual unbroken gauge group,
which is precisely the image of the diffeomorphism group
under the homomorphism defined by the fixed soldering form.
To see this, just write the transformation (\ref{transfsoldering})
with $\theta'=\theta$ and think of it as an equation for $\Lambda$.
Its solution is given by (\ref{injection}).
If we consider the torsion and nonmetricity in this gauge, 
we observe that the nonmetricity is still the covariant derivative 
of the metric, but the torsion becomes just an algebraic object:
\begin{equation}
\label{torsionmetric}
\Theta_\mu{}^a{}_\nu=A_\mu{}^a{}_\nu-A_\nu{}^a{}_\mu\,.
\end{equation}

If we consider the transformation (\ref{transfmetric})
we see that it is always possible to go to a gauge where
\begin{equation}
\label{vierbeingauge}
\gamma_{ab}=\delta_{ab}
\end{equation}
Looking at equation (\ref{pullbackmetric}) we see that in this gauge
(and only in this gauge) the soldering form can be interpreted as
the vierbein. For this reason we shall call this the vierbein gauge.
It leaves a residual unbroken gauge group,
which consists of orthogonal automorphisms, namely diffeomorphisms
and local transformations of the bases such that $\Lambda^a{}_b$
is an orthogonal transformation.
If we consider the torsion and nonmetricity in this gauge, 
we observe that the torsion is the covariant exterior derivative 
of the vierbein, but the nonmetricity becomes just an algebraic object:
\begin{equation}
\label{nonmetricityvierbein}
\Delta_{\lambda ab}=A_{\lambda ab}+A_{\lambda ba}
\end{equation}

From this discussion the following points emerge.
The fields $\gamma_{ab}$ and $\theta^a{}_\mu$
play the same role as the Goldstone bosons $\varphi^\alpha$ played in our discussion
of the Higgsless Higgs phenomenon.
I have already mentioned that $\gamma$ {\it  are} the Goldstone bosons that arise
when a global symmetry group $GL(N)$ is broken to $O(N)$
by the choice of a metric.
This is slightly less obvious in the case of the soldering form, 
mainly because it is not a scalar.
It is true also of the soldering form that its configuration space is a coset,
but this can only be seen at the level of infinite dimensional groups:
it is the quotient of the group of all automorphisms of $E$ by the
diffeomorphisms of $M$.
By a slight abuse of language from now on I will refer also to 
$\theta^a{}_\mu$ as a ``Goldstone boson'', so gravity is seen to be a
theory with two Goldstone bosons. 
The metric gauge and the vierbein gauge are analogs of the unitary
gauge, where each one of  the Goldstone bosons in turn takes a fixed value.
However, the gauge group is not big enough to fix both simultaneously. 
So in each one of the unitary gauges, the other Goldstone boson 
(either the metric or the vierbein) remains dynamical,
and it describes the geometry of spacetime.
Finally, choosing either one of the unitary gauges (which is the
standard procedure) hides the nature of torsion and/or
nonmetricity as the covariant derivatives of a field, since one
of them becomes an algebraic combination of components of the connections.
\smallskip

When gauge fields interact with Goldstone bosons we expect
the Higgs phenomenon to occur.
So is there a Higgs phenomenon in the case of gravity?
Do we see a connection becoming massive?
A first hint comes from looking at the so--called Palatini action
\begin{equation}
\label{palatini}
S_P(A,\gamma,\theta)={1\over16\pi G}\int d^4x\ \sqrt{|\det g|}\
\theta_a{}^\mu\theta^b{}_\rho\, g^{\rho\nu} F_{\mu\nu}{}^a{}_b\ ,
\end{equation}
where we have abridged $\theta^{-1}{}_a{}^\mu=\theta_a{}^\mu$.
If we assume that the vacuum of the theory is flat space,
$A=0$, $\theta=\mathbf{1}$, $\gamma=\mathbf{1}$,
and expand (\ref{palatini}) around the vacuum we see that it contains a term
\begin{equation}
\label{palatinimass}
{1\over16\pi G}\int d^4x\,
\delta{}_a{}^\mu\delta^{b\nu} 
(A_{\mu}{}^a{}_c A_{\nu}{}^c{}_b-A_{\nu}{}^a{}_c A_{\mu}{}^c{}_b)
\end{equation}
which is essentially a Planck mass for some components of the connection.
This is suggestive, but there are some important differences.
First, the Palatini action is very different from the kind of action that is
used in models of particle physics: it is not a Yang Mills action, because
it is of first order in the curvature, and it is not the covariant derivative
of some other field. It is obtained by contracting one of the form indices of
the curvature with one of the Lie algebra indices, and this is something
that one can only do with the soldering form.
Second, when a field is massive we normally expect it to vanish at very low
energies, in this instance much below the Planck mass,
but we know that in nontrivial solutions of Einstein's equations the connection
is far from being flat.

In order to properly understand what is going on, 
it is convenient to change variables.
let us recall that given $\theta$, $\gamma$, 
there is a unique connection $\bar A$, 
called the Levi Civita connection, such that $\Theta=0$ and $\Delta=0$.
Its components are
$$
\bar A={1\over2}\bigl(
{\theta}_c{}^\lambda\, \partial_\lambda\kappa_{ab}+
{\theta}_a{}^\lambda\, \partial_\lambda\kappa_{bc}-
{\theta}_b{}^\lambda\, \partial_\lambda\kappa_{ac}\bigr)
+{1\over2}\bigl(C_{abc}+C_{bac}-C_{cab}\bigr)
$$
where 
$C_{abc}=\kappa_{ad}\, \theta^d{}_\lambda\bigl({\theta}_b{}^\mu\,
\partial_\mu{\theta}_c{}^\lambda-
{\theta^{-1}}_c{}^\mu\, \partial_\mu{\theta^{-1}}_b{}^\lambda\bigr)$.
Any connection $A$ can be split uniquely as 
\begin{equation}
\label{split}
A=\bar A+\Phi\ .
\end{equation}
Then 
\begin{equation}
S(A,\gamma,\theta)=S(\bar A(\gamma,\theta)+\Phi,\gamma,\theta)=S'(\Phi,\gamma,\theta)\,.
\end{equation}

Let us now reconsider Einstein's theory from this point of view.
In first order formalism, the normal choice for the action is (\ref{palatini}).
A particle physicist should naturally ask: since the theory contains two Goldstone bosons,
where are their kinetic terms?
They have not been considered so far, so let us add them to the action.
The kinetic terms must contain the squares of the covariant derivatives
of the Goldstone bosons, {\it i.e.} torsion and nonmetricity:
$$
S(A,\gamma,\theta)=S_P(A,\gamma,\theta)+S_{\rm m}(A,\gamma,\theta)
$$
where
$$
S_{\rm m}%(\theta,\kappa,A)
\!=\frac{1}{16\pi G}\int \! d^4x \sqrt{|\det g|}
\left[A^\mu{}_a{}^{\nu\rho}{}_b{}^\sigma
\Theta_\mu{}^a{}_\nu\Theta_\rho{}^b{}_\sigma 
\! +\! B^{\mu ab\nu cd}\Delta_{\mu ab}\Delta_{\nu cd}
\!+\! C^\mu{}_a{}^{\nu\rho cd}\Theta_\mu{}^a{}_\nu\Delta_{\rho cd}\right]
$$
The tensors $A$, $B$, $C$ are combinations of $\gamma$ and $\theta$
which comprise the most general way of contracting six indices,
with arbitrary coefficients.
Two latin indices can be contracted with $\gamma$, two greek indices can be 
contracted with $g$
and a latin and a greek index can be contracted with $\theta$ or its inverse.
The action $S_{\rm m}$ must have a prefactor with dimension of mass squared,
which we can take to be $1/(16\pi G)$,
and contains several arbitrary dimensionless parameters
which we assume to be of order one
\footnote{It is in principle possible that some of the coefficients
in the action $S_{\rm m}$ are much smaller than one, so that some components
of $\Phi$ have masses which are much lower than the Planck mass.
It would be worthwile studying phenomenological implications of this scenario.}.
We can now insert (\ref{split}) into (\ref{torsion}) and (\ref{nonmetricity}) to obtain
the following formulae
\begin{eqnarray}
\Theta_\mu{}^a{}_\nu&=&\ 
\Phi_\mu{}^a{}_b\theta^b{}_\nu-\Phi_\nu{}^a{}_b\theta^b{}_\mu\nonumber\\
\Delta_{\mu ab}&=&\ 
\Phi_\mu{}^c{}_b\kappa_{ca}+\Phi_\mu{}^c{}_a\kappa_{cb}\nonumber\\
F_{\mu\nu}{}^a{}_b&=&\, \bar F_{\mu\nu}{}^a{}_b+\bar\nabla_\mu\Phi_\nu{}^a{}_b-
\bar\nabla_\nu\Phi_\mu{}^a{}_b+\Phi_\mu{}^a{}_c\,\Phi_\nu{}^c{}_b-
\Phi_\nu{}^a{}_c\,\phi_\mu{}^c{}_b\nonumber
\end{eqnarray}
\smallskip
and then rewrite the action, up to a total derivative, as
$$
S(A,\gamma,\theta)=S(\bar A+\Phi,\gamma,\theta)=S_H(\gamma,\theta)
+S_Q(\Phi,\gamma,\theta)
$$
where $S_H$ is the Hilbert action (which is identical to $S_P$ except that the
curvature of $A$ is replaced by the curvature of $\bar A$), and
$$
S_Q(\Phi,\gamma,\theta)
=\frac{1}{16\pi G}\int d^4x\ \sqrt{|\det g|}\  
Q^\mu{}_a{}^{b\nu}{}_c{}^d\, \Phi_\mu{}^a{}_b\, \Phi_\nu{}^c{}_d\ .
$$
The quadratic form $Q$ has the same general structure as $A$, $B$ and $C$,
with a prefactor $1/(16\pi G)$ and other dimensionless coefficients
of order unity, depending linearly on the coefficients of $A$, $B$, $C$.

We now see that the action we are considering depends on $\Phi$ only through the mass term.
So if the quadratic form $Q$ is nondegenerate, as will generally be the case,
the equation of motion of $\Phi$ will be simply $\Phi=0$, i.e. $A=\bar A$.
This explains why the connection does not necessarily vanish at low energy:
the correct statement is not that $A$ that is massive, 
but that the deviation of $A$ from $\bar A$ is.
We have seen that in any gauge there is always at least one dynamical Goldstone boson,
and its equations of motion can have nontrivial solutions.
Then, the connection $\bar A$ is nontrivial and therefore also the solution for $A$
is nontrivial.

Now the action $S_P+S_{\rm m}$ looks already a little more similar to the one that is
used in particle physics: the new terms $S_{\rm m}$ are the obvious kinetic terms for the
Goldstone bosons.
It is important to observe that if, as I will argue below, general relativity is
regarded as an effective quantum field theory, then at a fixed order of the momentum expansion
it would be {\it inconsistent} to leave out certain terms from the action, because quantum
corrections will generate them.
The Palatini action contains terms without derivatives of $\Phi$ and so does $S_{\rm m}$.
Therefore if we have the Palatini term in the action we {\rm must} also include
the terms $S_{\rm m}$.
Doing so does have a small but nontrivial effect on the dynamics:
if we had not added kinetic terms for the Goldstone bosons,
the quadratic form $Q$ would be degenerate and one would {\it not} get $A=\bar A$ as an
equation of motion.
The corresponding quadratic form is the one that appears in (\ref{palatinimass}) and one sees
that it vanishes identically on tensors of the form $\Phi_{\mu ab}=\xi_\mu\delta_{ab}$
(this is known as ``projective invariance'' of the Palatini action).
This is why in textbook formulations of Einstein's theory
one usually has to impose a priori the symmetry of the connection on lower
indices (vanishing torsion) and obtains the condition of metricity
from the equations of motion.
Or alternatively one can impose metricity and obtain zero torsion from the
equations of motion, but one cannot obtain both simultaneously.
However if one adds the kinetic terms for the Goldstone bosons
the quadratic form $Q$ becomes generically nondegenerate, and then
one gets both $\Delta=0$ and $\Theta=0$ from the equations of motion.

The natural next step is to consider also terms with two derivatives of the connection.
When terms quadratic in the curvature tensor are added,
as is most natural in view of the analogy with other gauge theories,
the equations of motion for $\Phi$ is no longer simply $\Phi=0$.
Rather, $\Phi$ becomes a propagating degree of freedom.
However it is still true that $\Phi$ has Planck mass,
so when one studies the theory at very low energies,
as we can only do, it is always a very good approximation to set $\Phi=0$.
But now recall that this is equivalent to setting $\Delta=0$ and $\Theta=0$,
and $\Delta$ and $\Theta$ are the covariant derivatives of
the Goldstone bosons.
So these conditions are the exact analog, in the case of gravity,
of the condition (\ref{lowenergy}).

This discussion sheds light on an otherwise baffling aspect of general relativity:
why does one impose that the connection be metric and that torsion vanishes?
It is clear that the connection plays a very important role in general relativity,
so why is it not allowed to have an independent dynamics?
The reason is that if we allow it to have an independent dynamics, 
then a gravitational Higgs phenomenon makes it massive,
and the natural mass is so large that at low energy we can effectively
pretend that the connection is not dynamical.
This is almost exactly the same as studying weak interactions at energies much lower than
the Fermi scale: there is a connection, but it is so massive that we cannot
excite it. Then the covariant derivative of the Goldstone bosons is zero.
The only reason why we do not usually view weak interactions this way
is that the Goldstone boson itself can be made constant by a gauge transformation,
and then the statement $D\varphi^\alpha=0$
is equivalent to saying that the connection is zero.
But in gravity one cannot make both Goldstone bosons simultaneously constant,
so there is always one left that can assume nontrivial configurations.

The general kinematical framework described here
would not help in solving the typical problems that one encounters in
general relativity, like finding the trajectory of a spacecraft or describing
gravitational collapse.
Its use would unnecessarily complicate matters, and for such applications
the familiar metric gauge is much more convenient.
In practice, it is useful if we want to understand certain formal aspects
of the theory. For example, it has been used to solve a puzzle
regarding the quantization of the Chern Simons term in three dimensional gravity \cite{tmg},
to understand the origin of the Bardeen Zumino anomaly counterterm \cite{anomaly},
and it is necessary to properly discuss the transformation of spinors under diffeomorphisms
\cite{dabrowski}.
In string theory, it has been used to give a linear realization of duality \cite{siegel}.
But its most important application may be in the understanding that there is
a Higgs phenomenon occurring in gravity.
In particle physics the Higgs phenomenon is used in the construction of unified theories:
Could the same be true in the case of gravity?

\section{GraviGUTs}

There have been many attempts to construct unified theories,
from Weyl's scale invariant theory \cite{weyl} to Kaluza and Klein's five dimensional theory \cite{kk},
later extended to nonabelian gauge theories \cite{kerner}, to superstrings.
Einstein famously spent the last part of his life in the unfruitful search for such a theory. 
See \cite{goenner} for a review of many such attempts.
Each of these theories achieves ``unification'' in a different way.
In this section I will use the word ``unification'' in the strict sense in which
it is used in particle physics, and I will discuss the possibility of 
achieving a unification of gravity with the other interactions in this sense.

A somewhat simplified description of a unified theory is as follows.
One has two gauge theories with gauge groups $G_A$ and $G_B$,
describing, say, ``A'' and ``B'' interactions.
To construct a unified theory one has to choose a group $G$ containing $G_A$ and $G_B$
as commuting subgroups, and then find an order parameter whose VEV will 
pick the two subgroups $G_A$ and $G_A$ inside $G$ and give mass (at least)
to the components of the gauge field that are not in $G_A\times G_B$.
In doing so, the VEV dynamically separates the ``A'' and ``B'' interactions, 
which in the starting theory are undifferentiated.
We would like to apply this same methodology also to gravity.
Since the term ``Grand Unification'' has already been used for the unified theories
of the weak and strong interactions, we will call ``Gravitational Grand Unified Theories''
or ``GraviGUTs'', those that contain also gravity.

Nobody has yet constructed a unified theory of gravity along these lines,
but there are hints that this may be possible.
I will describe here a few of them.
First the kinematics.
To construct a GraviGUT
one would assume that the fibers of the vectorbundle $E$ have dimension $N>4$, 
while the base manifold $M$ remains  four dimensional.
We would start therefore from a gauge theory for the group $GL(N)$.
For the metric $\gamma$, we assume that it is nondegenerate, with a given signature.
The soldering form cannot be assumed to be an isomorphism anymore;
the strongest condition we can require is that it has maximal rank $4$,
i.e. in geometrical terms that every tangent space $T_xM$ can be regarded as 
a linear subspace of the internal space $E_x$.

Then it can be seen that there exists an extended metric gauge where
\begin{equation}
\theta=\left[
\begin{array}{c}
\mathbf{1}_4\\ 
0\
\end{array}
\right]
\qquad,\qquad
\gamma=\left[
\begin{array}{cc}
g & 0 \\
0 & \mathbf{1}_{N-4}
\end{array}
\right]
\end{equation}
The connection can be split between $TM$ and the orthogonal complement:
\begin{equation}
A_\lambda=\left[
\begin{array}{cc}
A_\lambda^{(4)} & H_\lambda \\
K_\lambda & A_\lambda^{(N-4)}
\end{array}
\right]
\end{equation}
where $A_\lambda^{(4)}$ defines a connection in $TM$, $A_\lambda^{(N-4)}$ is a purely internal
Yang--Mills connection and $H$ and $K$ are fields mixing the internal and spacetime transformations.
As before, terms quadratic in torsion and nonmetricity generate masses for the connection,
more precisely for $A_\lambda^{(4)}$, $H$, $K$ and for the symmetric components of $A_\lambda^{(N-4)}$,
so that only the antisymmetric components of $A_\lambda^{(N-4)}$ remain massless \cite{percacci}.
These can be regarded as an $SO(N-4)$ Yang-Mills field.

It is most natural to take $N=14$, in which case the unbroken group is $SO(10)$,
which is already a well studied GUT.
One fact that remains unexplained in $SO(10)$ GUT is that the spinor of $SO(10)$ is also
a spinor of the Lorentz group.
We can gain new insight into this if we look at an underlying GraviGUTs.
For simplicity we can assume here that the connection is metric.
The spinor representations depend on the signature, so this is a subject where it is not
enough to consider the positive definite case.
If we consider groups $SO(p,q)$ with $p+q=14$ and containing $SO(10)\times SO(1,3)$ as a subgroup,
there are only two possibilities: $SO(3,11)$ and $SO(1,13)$.
The group $SO(3,11)$ has a real, 64-dimensional Majorana-Weyl representation.
When viewed as a representation of the subgroup $SO(3,1)\times SO(10)$,
it is a spinor of Lorentz and a spinor of $SO(10)$, and therefore it can be used
to describe a single fermionic family.
Similarly the group $SO(1,13)$ has a complex, 64-dimensional Weyl representation.
Under the subgroup $SO(1,3)\times SO(10)$ such a representation decomposes into
$$
\bf{64}=\bf{2} \times \bf{16} \oplus \mathrm{\bar 2} \times \mathrm{\bar {16}}
$$
These two representations are equivalent, so the Weyl spinor can be used to
describe two fermionic families.
The fact that the known fermions are spinors of Lorentz and spinors of $SO(10)$
can be seen as a hint in favor of this GraviGUT.

The gravitational Higgs phenomenon described above is of the Higgsless type:
the fields $\theta$ and $\gamma$, which a priori are tensorial objects,
are assumed to satisfy the nonlinear constraints on their rank and eigenvalues.
These constraints precisely say that $\theta$ and $\gamma$ must belong to
a single orbit of the gauge group.
In view of the fact that the introduction of the radial (Higgs) mode
makes the theory UV complete, one could speculate that
relaxing the nonlinear constraints on $\theta$ and $\gamma$
could improve the UV behaviour of gravity.
The central issue then becomes: where do the conditions 
$\mathrm{\det}\theta\not=0$ and the signature of $\gamma$ come from?
Or equivalently, why is gravity in the Higgs phase instead of being in
the symmetric phase?
One encounters here a new conceptual obstacle in the construction of a GraviGUT: 
in the symmetric phase the soldering form would vanish, and so would the metric. 
In other words, the symmetric phase of a GraviGUT is a topological phase, 
and one simply does not have all the tools that are available in ordinary GUTs.

This difficulty manifests itself in practice when one wants to write
down a dynamics for a GraviGUT.
Ideally, to mimick what we do in ordinary GUTs, one would like to be able to
write down a Lagrangian which is invariant under the unifying group $G$;
the difference between the gravitational and nongravitational interactions
should be due only to the VEV of the order parameter.
In ordinary GUTs, one can choose between different VEVs, and hence between
different phases of the theory, by tuning a few parameters in the Higgs potential.
It is not at all clear that one can do the same with gravity:
a potential is a term in a Lagrangian not involving derivatives
of the field, and the only covariant potential for gravity is the cosmological constant.
I am aware of two possible ways out.
One is to insist on constructing a nontrivial potential for the order parameters. 
This however requires using a second metric, which is somewhat unconventional
\footnote{A second metric also appears in application of functional renormalization group, 
as we shall see in sect. 5.}.
In \cite{percacci} I proposed using a bootstrap procedure, where one chooses 
a background metric, later to be identified of as the VEV of the metric, 
calculates the VEV of the metric in the chosen background and finally checks that the VEV
coincides with the background metric.
The other possibility is closer to the work that has been done on topological field theories:
different phases would appear as different solutions of the dynamical equations,
but there would not be a potential to select one as being energetically favored over another.

I will not discuss further the dynamics of the bosonic sector.
I refer to \cite{percacci} for some work along the first line, and to
\cite{graviweak,smolin} for work along the second line.
Instead, I will describe the fermionic sector of the $SO(3,11)$ GraviGUT mentioned above
\cite{gravigut}.
We start from the Clifford algebra of $SO(3,11)$, generated by gamma matrices
$\gamma^a$ (latin indices $a,b$ now run from $1$ to $14$), satifying 
$\{\gamma^a,\gamma^b\}=2\eta^{ab}$.
The $SO(3,11)$ covariant derivative acting on Majorana-Weyl spinors is
\begin{equation}
D_\mu\psi_{L+}
=\left(\partial_\mu+\frac{1}{2}A_\mu^{ab}\Sigma_{L\,ab}^{(3,11)}\right)\psi_{L+}
\end{equation}
where $\Sigma_{ab}^{(3,11)}={\tiny \frac{1}{4}}[\gamma^a,\gamma^b]$
are the generators of $SO(3,11)$ and $\Sigma_{L\,ab}^{(3,11)}$ 
their restriction to the Majorana-Weyl representation.
We also define the covariant differential $D$, mapping spinors to spinor-valued
one forms: $D\psi_{L+}=D_\mu\psi_{L+}dx^\mu$.
There is an intertwiner $A$ mapping the spinor representation to its 
hermitian conjugate: $\Sigma_{ab}^\dagger A=-A \Sigma_{ab}$.
Therefore the quadratic form
\begin{equation}
\label{eq:quad}
\psi_{L+}^\dagger (A\gamma^i)_LD\psi_{L+}
\end{equation}
is manifestly a vector under $SO(10)$ and a one form under diffeomorphisms. 
Then, to construct an $SO(10)$-invariant action, we introduce an auxiliary field
$\phi_{abcd}$ transforming as a totally antisymmetric tensor. The
action is
\begin{equation}
\label{eq:action}
\mathcal{S}=\int\psi_{L+}^\dagger (A\gamma^a)_LD\psi_{L+}\,\wedge\theta^b\wedge\theta^c\wedge\theta^d \,\phi_{abcd} \,.
\end{equation}
 
The breaking of the $SO(3,11)$ group to $SO(10)$
is induced by the VEV of two fields: the soldering one-form
$\theta^a{}_\mu$ and the four-index antisymmetric field $\phi_{abcd}$. We
assume that the VEV of $\phi_{abcd}$ is $\epsilon_{mnrs}$, the
standard four index antisymmetric symbol, in the Lorentz subspace 
(spanned by indices $m,n=1,2,3,4$), and
zero otherwise.\footnote{The field $\phi_{abcd}$
also appears in Plebanski reformulations of General Relativity,
where the soldering form is traded for a two form field,
which is equivalent to $\theta$ on shell \cite{smolin}.} 
The VEV of the soldering form on the other
hand has maximal rank (four) and is also nonvanishing only in the
Lorentz subspace:
\begin{equation}
\left\{\begin{array}{l}
\phi_{mnrs}=\epsilon_{mnrs}\\
\phi_{abcd}=0 \quad\mathrm{otherwise}
\end{array}
\right.
\quad
\left\{\begin{array}{l}
\theta^m{}_\mu=M e^m{}_\mu\\
\theta^a{}_\mu=0 \quad\mathrm{otherwise}
\end{array}
\right.
\label{eq:VEVs}
\end{equation}
where $e^m{}_\mu$ is a vierbein, corresponding to some solution of the
gravitational field equations which we need not specify in this
discussion (below we will choose $e^m{}_\mu=\delta^m_\mu$) and $M$ can
be identified with the Planck mass.

Then the action for fluctuations around this VEV reduces to the standard 
action for a single $SO(10)$ family in flat space:
\begin{equation}
%\label{eq:action}
\int d^4x\,\eta^\dagger\sigma^\mu \nabla_\mu\eta \,,
\end{equation}
where now
$\nabla_\mu=D^{(10)}_\mu=\partial_\mu+\frac{1}{2}A_{\mu\,(10)}^{ab}\Sigma^{(10)}_{ab}+\frac{1}{2}A_{\mu\,(3,1)}^{mn}\Sigma^{(3,1)}_{mn}$
is the $SO(10)$ covariant derivative. Note that this action contains
the standard kinetic term of the fermions, and the interaction with
the $SO(10)$ gauge fields, which at this stage can still be assumed to
be massless. As discussed above, the Lorentz connection
$A_{\mu\,(3,1)}^{mn}$ in the covariant derivative can be assumed to be
the Levi-Civita connection derived from the vierbein. Its fluctuations
around this VEV are also present but have a mass of the order of the
Planck mass and are negligible at low energies.

To summarize, it looks like the fermionic sector of a realistic GraviGUT can be
constructed without encountering major difficulties.
The bosonic sector probably poses greater challenges.
In particular, as I have already mentioned before, there are deep issues
concerning the dynamical mechanism that generates the necessary VEVs.
This is not unexpected, since to date the symmetry breaking mechanism 
is still somewhat unclear even in the case of the standard model.
But probably the main difficulty in the construction of a GraviGUT
is that it has to be a quantum field theory of gravity.
In the next lecture I will describe progress on this issue.

\section{Asymptotic Safety}

It is now well understood that gravity can be treated as an {\it effective quantum field theory},
exactly in the same way as the NLSM \cite{donoghue,burgess}.
If one applies perturbation theory to General Relativity,
one finds that it is an expansion in the parameter $\tilde G=G k^2$, 
where $k$ is a characteristic momentum scale of the problem.
For example, in a hypothetical graviton scattering experiment, it could be related to
one of the Mandel'stam parameters.
At all accessible energies $\tilde G$ is extremely small, 
so tree level perturbation theory works well.
One can in principle compute loop corrections
putting a cutoff at the Planck scale,
but they are unmeasurably small at available energies.
So every experimental evidence for General Relativity is also
evidence for this effective theory of gravity.

The difficulties of quantum gravity only become apparent if one tries to
reach the Planck scale, or even more dramatically if one tries to remove
the UV cutoff. One can actually distinguish two types of issues.
The first is that the coupling $\tilde G$ diverges in the infinite cutoff limit.
This would lead to  unacceptable divergences in physically measurable quantities.
The second issue is that at each order of the expansion new divergences appear,
such that they cannot be reabsorbed into renormalizations of a finite number of couplings
\cite{thooft,sagnotti}.
There is no logical inconsistency in renormalizing an infinite number of terms,
but then the theory loses its predictive power, because all the counterterms
have to be fixed by experiment.

This state of affairs has led many physicists to doubt the capacity of 
quantum field theory to properly
describe gravity at high energies.
There is however a logical possibility that has not been excluded so far, namely that
the theory can be made sense of using nonperturbative methods.
Loop Quantum Gravity is a nonperturbative approach based on canonical methods
\cite{ashtekar,lqg}.
Regge calculus and dynamical triangulations provide discrete approximations
similar in spirit to lattice QFT \cite{hamber,loll}.
Here I will describe another approach that uses continuum, covariant QFT methods
and was described in \cite{weinberg}: it goes under the name of ``Asymptotic Safety''.

To avoid the two classes of problems that were mentioned above, one could
require that the following situation be realized.
First, the strength of the coupling must cease to grow at high energies.
This can happen if we take into account that Newton's constant,
like every other coupling constant in the action, 
will be subject to Renormalization Group (RG) flow.
The quantity $\tilde G(k)$, which naively grows linearly with $k$,
is really $\tilde G(k)=G(k) k^2$,
where Newton's constant is momentum--dependent and, for a process occurring at
energy $k$, will have to be evaluated at the scale $k$.
It is conceivable that when $k$ reaches the Planck scale, 
$G(k)$ will begin to scale like $k^{-2}$;
then $\tilde G(k)$, will stop growing and will tend to a constant.
This means that Newton's constant has a Fixed Point (FP).
%
%\footnote{In particle physics it is more common to discuss the RG flow
%of dimensionless couplings. A FP is then simply a point where the beta function of
%the coupling vanishes. When dimensionful couplings are involved, the definition
%of a FP is a point where the beta function of the coupling measured in cutoff units
%vanishes.}
%

More generally, we can write a Wilsonian, scale dependent effective action $\Gamma_k$
as a sum of operators $\calo$ constructed with the fields and their derivatives,
multiplied by scale dependent couplings $g$. 
In the spirit of effective field theories, we can assume that it admits
a derivative expansion
\begin{eqnarray}
\label{expansion}
\Gamma_k(g)&=&\sum_{n=0}^\infty\sum_i g^{(n)}_i {\calo}^{(n)}_i\nonumber\\
&=&\int d^4x\,\sqrt{g}
\left[ 2Z_g\Lambda-Z_{g}R
+\frac{1}{2\lambda}C^2
+\frac{1}{\xi}R^2
+\frac{1}{\rho}E
+\frac{1}{\tau}\nabla^2 R
+\ldots\right]
\end{eqnarray}
where ${\calo}^{(n)}_i$ contains $n$ derivatives of the metric and
$i$ is an additional index that labels terms with the same number of derivatives.
The first few terms in this expansion, containing up to four derivatives, are listed explicitly:
$C^2$ is the square of the Weyl tensor and 
$E=R_{\mu\nu\rho\sigma}R^{\mu\nu\rho\sigma}-4R_{\mu\nu}R^{\mu\nu}+R^2$ 
is ($32\pi^2$ times) the integrand of the Euler term, a total derivative in $d=4$. 
We define dimensionless quantities
\begin{equation}
\tilde g_i=k^{-d_i}g_i\ ,
\end{equation}
corresponding to the couplings $g_i$ measured in units of $k$.
Defining $t=\log\left(\frac{k}{k_0}\right)$ and $\beta_i=\partial_t g_i$, 
a gravitational FP is a point in the space of all couplings where
\begin{equation}
\tilde\beta_i\equiv \partial_t\tilde g_i=-d_i\tilde g_i+k^{-d_i}\beta_i
\end{equation}
all vanish simultaneously
\footnote{Strictly speaking one need to impose this condition only on the so--called
essential couplings, namely those that cannot be fixed by field redefinitions.}.
If the actual RG trajectory that describes our world tends to a FP, {\it i.e.} 
$\tilde g_i(k) \to \tilde g_{i*}$ when the RG parameter $t=\log\frac{k}{k_0}\to\infty$,
then the theory is safe from the type of divergences described above
\footnote{This does not mean that there will be no divergences at all:
the couplings $g_i^{(n)}$ with positive mass dimension will still diverge with powers of $k$,
but these powerlike divergences are harmless: what is physically important is that the
dimensionless couplings $\tilde g_i^{(n)}$ be finite.
So the overall behaviour of the theory will be under control;
for example, a cross section will behave like $k^{-2}$,
times a function of $k$ that tends to a constant.}.
The fixed point regime is characterized by the fact that
every dimensionful quantity will scale with $k$ exactly as required by
its canonical dimension.

If every trajectory in the space of all couplings had this good asymptotic behaviour,
then the initial conditions for the flow would be arbitrary; all the couplings
would have to be determined by comparison with experiment 
and the theory would again be as unpredictive as a nonrenormalizable theory.
Thus, to fix the second set of problems we have to require that only a
finite number of parameters is left free by the condition of having a good UV behaviour.
This will be the case if the UV critical surface, defined as the
locus of the points that flow towards the FP when $t\to\infty$, has finite dimension $d_{UV}$.
The requirement of a good asymptotic behaviour demands that the flow
starts on this surface, and therefore all but a finite number of couplings is
determined. 
Explicitly, we can choose for example the first $d_{UV}$ couplings as coordinates
in the critical surface; the values of these parameters at a given energy scale
will have to be determined by experiment, and all the others will then be fixed
by equations of the form
$\tilde g_k=\tilde g_k(\tilde g_1,\ldots,\tilde g_{d_{UV}})$.
In principle these relations could be turned into relations between physical
observables and therefore constitute predictions of the theory.

In practice one can determine the position of the FP and the
tangent space to the critical surface at the FP.
This can be done by studying the linearized flow equations
\begin{equation}
\partial_t y_i=M_{ij}y_j
\end{equation}
where $y_i=\tilde g_i-\tilde g_{i*}$ and
$M_{ij}={\partial \tilde\beta_i\over\partial\tilde g_j}\bigr|_*$.
Let $S$ be the linear transformation that diagonalizes $M$:
$S_{ik}^{-1}M_{k\ell}S_{\ell n}=\delta_{in}\lambda_n$.
The linearized RG equation for the variables $z_i=S_{ik}^{-1}y_k$ is
$\partial_t z_i=\lambda_i z_i$, where $\lambda_i$ are the eigenvalues of $M$,
so $z_i(t)=\exp(\lambda_i t)=\left(\frac{k}{k_0}\right)^{\lambda_i}$.
One also defines the ``critical exponents'' $\vartheta_i=-\lambda_i$.
The coordinates $z_i$ that correspond to negative eigenvalues (positive critical exponents)
are attracted to the FP and are called relevant.
Those corresponding to positive eigenvalues are repelled by it and are called irrelevant.
Therefore, the tangent space to the critical surface at the FP is the space spanned
by the eigenvectors with negative eigenvalue.
In particular, the dimension of the critical surface is equal to the number
of negative eigenvalues of the matrix $M$.

A theory with a FP and a finite--dimensional UV critical surface is said to be
Asymptotically Safe.
An example of an asymptotically safe theory is QCD. 
In this case the FP is the Gaussian FP (the free theory). 
Because the beta functions arise from loop effects, $\beta_i$ vanish at the Gaussian FP.
The matrix $M$ is given by $M_{ij}=-d_i\delta_{ij}$, so the UV--attractive 
(relevant) couplings are those that have positive mass dimension.
Near the origin, the UV critical surface is simply the space spanned
by the renormalizable couplings.
This example shows that symptotic safety at a Gaussian FP is equivalent to
the statement that the theory is perturbatively renormalizable and asymptotically free.
So we also see that asymptotic safety is a generalization of this good behavior,
where we replace the Gaussian FP by a nontrivial one.

Perturbation theory is defined in a neighborhood of the Gaussian FP,
so if the theory tends to a nontrivial FP in principle we lose the ability
to perform arbitrarily accurate predictions.
In practice, however, if the nontrivial FP is not too far from the Gaussian one,
perturbation theory may still be of use even for quantitative calculations.
%For example some of the best estimates of the properties of the Wilson-Fischer FP
%of three-dimensional scalar theory is by using high orders of perturbation theory.

We know that gravity is not perturbatively renormalizable.
However, it could still be asymptotically safe.
There is no logical reason to exclude such behaviour.
In fact, all the evidence collected so far supports this hypothesis, as I will describe next.

\subsection{One loop, $\varepsilon$ expansion and large $N$}

The first evidence that gravity could be asymptotically safe 
came from the expansion around two dimensions.
In $2+\epsilon$ dimensions Newton's constant $G$ has mass dimension $-\epsilon$;
defining $\tilde G=G k^\epsilon$, its beta function is \cite{weinberg,epsilon}:
\begin{equation}
\label{betaepsilon}
\partial_t\tilde G=\epsilon\tilde G-\frac{38}{3}\tilde G^2\ .
\end{equation}
The beta function is plotted, with $\epsilon=2$ in Fig.1.
It has an infrared--attractive FP at zero, and an UV--attractive FP
at $\tilde G=3\epsilon/38$.
\begin{figure}[htp]
\centering
\includegraphics{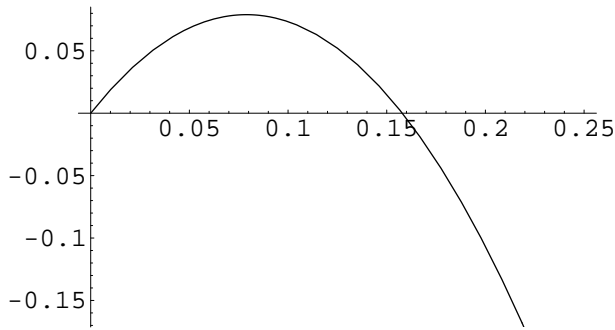}
\caption{The beta function for Newton's constant in the $\epsilon$ expansion, for $\epsilon=2$.}
\end{figure}
If the $\epsilon$ expansion was reliable up to $\epsilon=2$ then we
would have a nontrivial FP in four dimensional gravity.
Unfortunately there is no a priori reason to believe that the $\epsilon$
expansion is a reliable guide for such large values of $\epsilon$,
so this evidence is not very strong.

One would like to be able to compute directly in four dimensions.
If we follow a logical rather than a historical order, the next
step would be a one loop calculation of the beta function for Newton's
constant directly in four dimensions.
This can be extracted from \cite{bjerrum}.
They identify a subset of graphs which can be interpreted as giving
a distance--dependence of Newton's constant:
$$
G(r)=G_0\left[1-\frac{167}{30\pi}\frac{G_0}{r^2}\right]\ ,
$$
where $r$ is the distance between two gravitating point particles.
If we identify $k=1/ar$, with $a$ a constant of order one,
this would correspond to a beta function
\begin{equation}
\label{betabjerrum}
\beta_{\tilde G}=2\tilde G-a^2\frac{167}{15\pi}\tilde G^2\ .
\end{equation}
This beta function has the same form as (\ref{betaepsilon})
in four dimensions,
and, most important, the second term is again negative. This means
that the dimensionful Newton constant $G$ {\it decreases} towards
lower distances or higher energies, \ie\ gravity is {\it
antiscreening}. This is the behavior that is necessary for a FP to exist,
and indeed this beta function predicts a FP for
$\tilde G=\frac{30\pi}{167 a^2}$.
This calculation was based on perturbative methods and
since the FP occurs at a not very small value of $\tilde G$,
it is again not clear that one can trust the result.
What we can say with confidence is that the onset of the running of $G$
has the right sign.
Clearly in order to make progress on this issue we need different tools.

Another approximation that can yield nonperturbative information is
the $1/N$ expansion \cite{smolinn,largen}.
In gravity, this consists in assuming that the number of matter fields
is very large.
Let us assume that there are $n_S$ scalars, $n_D$ Dirac, $n_M$ Maxwell
fields, all massless and minimally coupled to gravity, with $n_S$, $n_D$
and $n_M$ all of order $N$.
Matter loops contribute to the running of the gravitational couplings
and in the limit $N\to\infty$ they are dominant over the graviton contribution.
In the leading order of the approximation one simply drops the graviton terms.
\footnote{Note that this may be a good approximation in the real world.}
In four dimensions the beta functions have the form \cite{largen}
\begin{equation}
\partial_t \tilde g^{(n)}_i=(n-4)\tilde g^{(n)}_i+a^{(n)}_i
\end{equation}
where $a^{(n)}_i$ are constants, depending only on the number of matter fields.
The first few constants, corresponding to the operators written in (\ref{expansion}), are
\begin{eqnarray}
a^{(0)}&=&\frac{1}{32\pi^2}\left(n_S-4n_D+2n_M\right)\\
a^{(2)}&=&\frac{1}{96\pi^2}\left(n_S+2n_D-4n_M\right)\\
a^{(4)}_1&=&\frac{1}{2880\pi^2}\left(\frac{3}{2}n_S+9n_D+18n_M\right)\\
a^{(4)}_2&=&\frac{1}{2880\pi^2}\left(-\frac{1}{2}n_S-\frac{11}{2}n_D-31n_M\right)\\
a^{(4)}_3&=&\frac{1}{2880\pi^2}\frac{5}{2}n_S\\
a^{(4)}_4&=&\frac{1}{2880\pi^2}\left(6n_S+3n_D-18n_M\right)\\
\end{eqnarray}
From here one sees immediately that for all $n\not= 4$ there is a FP at
\begin{equation}
\tilde g^{(n)}_{i*}={1\over 4-n}a^{(n)}_i
\end{equation}
whereas for $n=4$ the couplings run logarithmically: 
$g^{(4)}_i(k)=g^{(4)}_i(k_0)+a^{(4)}_i\ln(k/k_0)$.
There follows that the couplings $\xi$, $\rho$ and $\tau$ in (\ref{expansion})
are asymptotically free.
Writing $g^{(0)}=\frac{2\Lambda}{16\pi G}$ and $g^{(2)}=\frac{1}{16\pi G}$
we find the following beta functions for the conventional
cosmological constant and Newton's constant:
\smallskip
\begin{equation}
\label{simpleflow}
\partial_t\tilde\Lambda=-2\tilde\Lambda +8\pi a^{(0)}\tilde G
+16\pi a^{(2)}\tilde G\tilde\Lambda\ ;\ \ \
\partial_t\tilde G=2\tilde G+16\pi a^{(2)}\tilde G^2
\end{equation}
\medskip
which have a FP at
\begin{equation}
\tilde G_*=\frac{12\pi}{-n_S-2 n_D+4 n_M}\ ;\ \ 
\tilde\Lambda_*=\frac{3}{4}\left(\frac{n_S-4 n_D+2 n_M}{-n_S-2 n_D+4 n_M}\right)\ .
\smallskip
\end{equation}
\smallskip
The qualitative shape of the flow of these variables is shown in Fig.2. (The FP occurs for positive
or negative $\tilde\Lambda$ depending on the difference between bosonic
and fermionic degrees of freedom).

\begin{figure}[htp]
\centering
\includegraphics{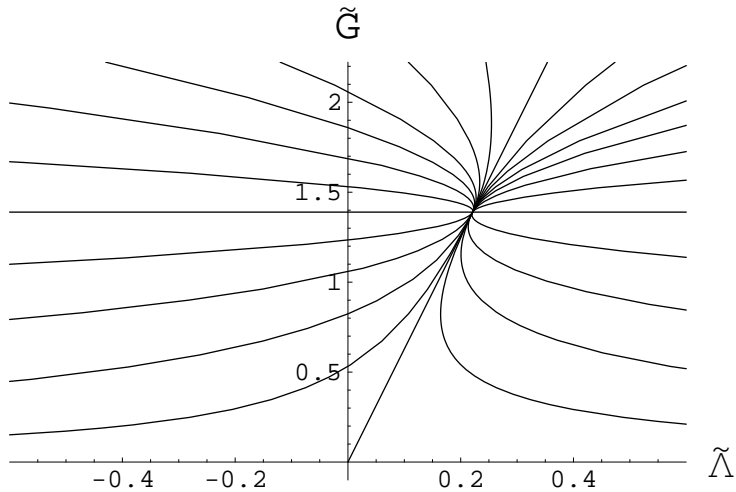}
\caption{The flow in the $\tilde\Lambda$--$\tilde G$ plane corresponding to eq. (5.23).}
\end{figure}

This approximation has the remarkable property that one proves the
existence of the FP for {\it all} the gravitational couplings
in the derivative expansion (\ref{expansion}).
Furthermore, using the ``optimized'' cutoff $R_k(z)=(k^2-z)\theta(k^2-z)$
\cite{optimized}, 
one finds $\tilde g^{(n)}_{i*}=0$ for $n\geq 3$.
%This suggests that by a change of variables 
%it may be possible to achieve a simple form of the FP action,
%with only finitely many terms.

It is possible to compute the one loop contribution of gravitons to the
beta functions of the terms listed in the second line of equation (\ref{expansion}),
taking into account the contribution of the four derivative terms.
This problem has a long history. 
It was proven that a generalization of Einstein's theory containing 
terms quadratic in the curvature tensor is renormalizable in flat space 
perturbation theory \cite{stelle}.
It was also established in a series of papers 
\cite{julve,fradkin,avramidi,shapiro} that the dimensionless
couplings $\xi$, $\rho$ and $\tau$ are asymptotically free.
The calculation was repeated in \cite{codello1} using the same gauge fixing
condition as the old papers, but using a momentum cutoff.
See also \cite{niedermaier2}.

It is customary to define $\frac{1}{\xi}=-\frac{\omega}{3\lambda}$
and $\frac{1}{\rho}=\frac{\theta}{\lambda}$; in this way $\lambda$ gives
the overall behaviour of the curvature squared terms while $\omega$ and
$\theta$ give the relative weight of the $R^2$, Weyl squared and
Euler terms. 
The beta functions of the dimensionless couplings are
\begin{eqnarray}
\beta_{\lambda} & =&  -\frac{1}{(4\pi)^{2}}\frac{133}{10}\lambda^{2}\ ,\\
\beta_{\omega} & =&  -\frac{1}{(4\pi)^{2}}\frac{25 + 1098\,\omega+ 200\,\omega^2}{60}\lambda\ ,\\
\beta_{\theta} & =&  \frac{1}{(4\pi)^{2}}\frac{7(56-171\,\theta)}{90}\lambda\ .
\end{eqnarray}
A FP occurs for $\omega(k)\to \omega_*\approx -0.0228$, $\theta(k)\to \theta_*\approx 0.327$,
and $\lambda$ gives asymptotic freedom for all curvature squared terms:
\begin{equation}
\lambda(k)=
\frac{\lambda_0}{1+\lambda_0\frac{1}{(4\pi)^{2}}\frac{133}{10}\log\left(\frac{k}{k_0}\right)}
\ .
\end{equation}
The beta functions of the cosmological constant and Newton's constant are
\begin{eqnarray}
\beta_{\tilde \Lambda} & =&
-2\tilde\Lambda
+\frac{1}{(4\pi)^{2}}\left[
\frac{1+20\omega^2}{256\pi\tilde G\omega^2}\lambda^2
+\frac{1+86\omega+40\omega^2}{12\omega}\lambda\tilde\Lambda\right]
-\frac{1+10\omega^2}{64\pi^2\omega}\lambda
+\frac{2\tilde G}{\pi}
-q(\omega)\tilde G \tilde\Lambda\nonumber\\
\beta_{\tilde G} & =&  2\tilde G
-\frac{1}{(4\pi)^{2}}\frac{3+26\omega-40\omega^2}{12\omega}\lambda\tilde G
-q(\omega) \tilde G^2\ ,
\end{eqnarray}
where $q(\omega)=(83+70\omega+8\omega^2)/18\pi$.
If we set $\lambda\to 0$ they reduce to
\begin{equation}
\label{qflow}
\partial_t\tilde\Lambda=
-2\tilde\Lambda
+\frac{2\tilde G}{\pi}
-q_*\tilde G \tilde\Lambda\ ;\ \ 
\partial_t\tilde G=  2\tilde G-q_* \tilde G^2\ ,
\end{equation}
where $q_*=q(\omega_*)\approx 1.440$.
These beta functions have the same form as in (\ref{simpleflow}), except for the
fact that the coefficients now depend on $\omega$
instead of the numbers of matter fields $n_A$.
The flow in the $\tilde\Lambda$--$\tilde G$ plane is shown in Fig.2.
In particular, for pure gravity the FP now occurs at
\begin{equation}
\tilde{\Lambda}_{*}=\frac{1}{\pi q_*}\approx 0.221\ ,\ \ \ \ \
\tilde {G}_{*}=\frac{2}{q_*}\approx 1.389\ .  
\end{equation}
The critical exponents are $-4$ and $-2$, and
the dimensionless couplings $\lambda$, $\xi$ and $\rho$
are marginal in this approximation.

\subsection{The Exact  Renormalization Group Equation}

As mentioned in the Introduction, most of the progress of the last ten years has come from applying
functional renormalization group methods to gravity.
The general idea of Wilson is that the functional integration should not be
performed in one single step covering all field fluctuations from the UV to the IR,
weighting all fluctuations with the same bare action,
but rather in a sequence of finite steps, updating the action at each step.
A concrete implementation of this idea that is easily amenable to explicit
calculations was given in 1993 by Wetterich \cite{wetterich}.
We begin from a formal functional integral 
\begin{equation}
e^{-W_k[J]}=\int (d\Phi) e^{-S(\Phi)+\Delta S_k(\Phi)+\int J\Phi}
\end{equation}
where $J$ is an external source and
$\Delta S_k(\Phi)=\frac{1}{2}\int d^4q \Phi(-q) R_k(q^2) \Phi(q)$.
The effect of the new term $\Delta S_k$ is simply to modify the (inverse)
propagator of the theory: it replaces $q^2$ by $P_k(q^2)=q^2+R_k(q^2)$.
The kernel $R_k(q^2)$ is chosen so as to suppress the propagation
of the modes with momenta $|q|\ll k^2$ but tends to zero for $|q|\gg k^2$
so that high momentum modes are unaffected.

One then defines a scale--dependent effective action functional $\Gamma_k(\Phi)$, 
as the Legendre transform of $W_k$,
minus the term $\Delta S_k$ that we introduced in the beginning:
\begin{equation}
\label{Gamma}
\Gamma_k[\Phi]=W_k[J]-\int J\Phi-\Delta S_k(\Phi)\ ,
\end{equation}
where $\Phi$ is now to be interpreted as a shorthand for $\langle\Phi\rangle$,
the variable conjugated to $J$.
If the functional integral is defined by an UV cutoff, then when $k$ tends to
this cutoff the average effective action is related by a nontrivial transformation
to the bare action \cite{manrique1}.
For $k\to 0$, $\Delta S_k\to 0$ and one
recovers the standard definition of the effective action (the generating function
of one--particle--irreducible Green functions).
It is not exactly the Wilsonian action but its definition is similar in spirit
and it is somewhat easier to work with.

If one evaluates this functional at one loop, it is 
\begin{equation}
\Gamma_k^{(1)}=\frac{1}{2}\STr \log\left({\delta^2 S\over\delta\Phi\delta\Phi}+R_k\right)
\end{equation}
and its scale dependence is given by
\begin{equation}
k\frac{d\Gamma_k}{dk}=
{1\over 2}
\STr\left({\delta^2 S\over\delta\Phi\delta\Phi}+R_k\right)^{-1}
k\frac{dR_k}{dk}\ .
\end{equation}
Here STr is a trace
that includes a factor $-1$ for fermionic fields and a factor $2$ for
complex fields. 
It can be shown that the ``renormalization group improvement'' of this equation,
which consists in replacing $S$ by $\Gamma_k$ in the r.h.s.,
leads actually to an exact equation often called the Exact Renormalization Group Equation
(ERGE) \cite{wetterich}:
\begin{equation}
\label{erge}
k\frac{d\Gamma_k}{dk}=
{1\over 2}
\STr\left({\delta^2 \Gamma_k\over\delta\Phi\delta\Phi}+R_k\right)^{-1}
k\frac{dR_k}{dk}\ .
\end{equation}

From (\ref{expansion}) one obtains
\begin{equation}
\partial_t \Gamma_k=\sum_{n=0}^\infty\sum_i \beta^{(n)}_i(k) {\cal O}^{(n)}_i
\end{equation}
so, if the operators $\calo_i^{(n)}$ form in some sense a complete set,
expanding the trace in (\ref{erge}) on this basis one can read off the beta
functions of all couplings.

It is important to observe that the last term in (\ref{erge}) suppresses the contribution
of high momentum modes so that the trace is ultraviolet finite: there is no need
to use any ultraviolet regularization.
In fact, once the equation has been derived, it is actually not necessary to
refer to the functional integral anymore.
The ERGE defines a flow in the space of all theories and if we start from any
point and we follow the flow in the limit $k\to 0$, then we find the
effective action, from which in principle we can derive everything we may
want to know about the theory.
Conversely, by following the flow towards higher energy we can establish
whether the theory has a FP with the desired properties.
If it does, this has to be taken as the initial point of the RG flow.
Thus, in this approach one does not make any {\it a priori} assumption about the
bare theory, except for the nature of the relevant degrees of freedom and its symmetries.
Instead, the starting point of the quantization will be determined as a result of
this study.

The application of this equation to gravity has been discussed first in \cite{reuter1}.
Since gravity is a gauge theory, one has to take into account the complications
due to the gauge fixing and ghost terms.
So far the best way to deal with these complications is to use the background field method.
Let $\bar g_{\mu\nu}$ be a fixed but otherwise arbitrary metric. We can write 
$g_{\mu\nu}=\bar g_{\mu\nu}+h_{\mu\nu}$. It is not implied that $h$ is small.
We choose a gauge--fixing condition of the form
\begin{equation}
S_{GF}(\bar g, h)=\int d^4x\sqrt{\bar g}\,\chi_{\mu}Y^{\mu\nu}\chi_{\nu}
\end{equation}
where $\chi_{\nu}=\nabla^{\mu}h_{\mu\nu}+\beta\nabla_{\nu}h$
and $Y$ is some operator, which in the simplest cases is just equal to $\bar g_{\mu\nu}$.
The standard formal manipulations in the path integral give rise to a ghost term
\begin{equation}
S_c=\int d^4x\sqrt{\bar g}\,\bar{c}_{\nu}(\Delta_{gh})_{\mu}^{\nu}c^{\mu}\ ,
\end{equation}
and, if $Y$ contains derivatives, also a ``third ghost'' term \cite{bos}
\begin{equation}
S_b=\frac{1}{2}\int d^4x\sqrt{\bar g}\,b_{\mu}Y^{\mu\nu}b_{\nu}\ .
\end{equation}
Also the cutoff term $\Delta S_k$ is written in terms of the background metric
\begin{equation}
\Delta S_k(\bar g)=\int d^4x\sqrt{\bar g}\,  
h_{\mu\nu}\bar g^{\mu\rho}\bar g^{\nu\sigma}R_k(\bar\Delta)h_{\rho\sigma}
\end{equation}
where $\bar\Delta$ is some differential operator constructed with the
background metric.

In this way one constructs a generating functional $W(j^{\mu\nu},\bar g_{\mu\nu})$
depending on sources that couple linearly to $h_{\mu\nu}$, and on the
background metric.
Applying the definition (\ref{Gamma}) one obtains a functional $\Gamma_k(h_{\mu\nu},\bar g_{\mu\nu})$
where $h$ is now a shorthand for $\langle h\rangle$, the Legendre conjugate of $j^{\mu\nu}$.
One can also think of $\Gamma_k$ as a functional of two metrics, namely
$\langle g_{\mu\nu}\rangle=\bar g_{\mu\nu}+\langle h_{\mu\nu}\rangle$
and the background metric.
In the limit $k\to 0$ this functional becomes the ordinary gravitational 
effective action in the background gauge.
The functional $\Gamma_k(g,\bar g)$ is invariant under simultaneous
coordinate transformations of $g$ and $\bar g$, the so--called background
gauge transformations.
We will restrict our attention to the functional
$\Gamma_k(g)=\Gamma_k(g,g)$ obtained by the identification of the background field
(which hitherto remaind completely unspecified) and the vacuum expectation value $g$.
By construction this functional has the same gauge invariance as the
original action and it contains the information about the familiar terms such
as the Einstein--Hilbert action. The functional $\Gamma_k(g,\bar g)$ 
contains in addition the information about the $k$--dependence of the gauge--fixing terms
and other genuinely bimetric terms in the action.
In the following we will ignore the RG flow of these terms.
The functional $\Gamma_k(g,\bar g)$ obeys an ERGE that has the same form
as in (\ref{erge}), where $\Phi$ stands for $h_{\mu\nu}$, $c$, $\bar c$ and $b$. 
We are now ready to discuss approximations schemes that allow the gravitational
beta functions to be extracted from this ERGE.

\subsection{Two derivative truncations}

A nonperturbative way of approximating the ERGE is to truncate the average effective action
to a finite number of terms, introducing them into the ERGE and reading off the beta functions.
Aside from the truncation, there is then no other approximation.
Let us consider first the Einstein--Hilbert truncation, which consists of retaining
only the terms linear in $R$ in the action. 
In the gauge $\alpha=Z$ and using the optimized cutoff the beta functions have
the following form
\begin{eqnarray}
\beta_{\tilde \Lambda}&=&
\frac{-2(1-2\tilde\Lambda)^2\tilde\Lambda
+\frac{36-41\tilde\Lambda+42\tilde\Lambda^2-600\tilde\Lambda^3}{72\pi}\tilde G
+\frac{467-572\tilde\Lambda}{288\pi^2}\tilde G^2}
{(1-2\tilde\Lambda)^2-\frac{29-9\tilde\Lambda}{72\pi}\tilde G}\\
\beta_{\tilde G}&=&  
\frac{2(1-2\tilde\Lambda)^2\tilde G
-\frac{373-654\tilde\Lambda+600\tilde\Lambda^2}{72\pi}\tilde G^2}
{(1-2\tilde\Lambda)^2-\frac{29-9\tilde\Lambda}{72\pi}\tilde G}
\end{eqnarray}

If one approximates the denominator by one and neglects $\tilde\Lambda$
the beta function of $\tilde G$ takes the one loop form (\ref{betabjerrum}).
The form of the flow is similar to the one obtained in the previous approximations,
except that the eigenvalues of the linearized flow are a complex conjugate pair
$\theta_0=\theta_1^*=-1.69\pm 2.49 i$
and therefore the approach to the FP follows spiralling trajectories.
The FP occurs at $\tilde \Lambda_*=0.171$ and $\tilde G_*=0.701$.
The flow is illustrated in Fig.3.
The FP was found in this truncation in \cite{souma}; 
its stability under changes of gauge 
and changes of cutoff has been discussed in \cite{lauscher1}.
It is also possible to follow the flow for varying spacetime dimensionality $d$
\cite{fischer}.
One interesting by--product of this calculation is that 
the derivative of $\tilde G_*$ with respect to $d$
at $d=2$ is $3/38$. The result of the $\epsilon$--expansion is thus vindicated:
the FP that is found in four dimension is indeed the continuation of the one
that is predicted by equation (\ref{betaepsilon}). 

\begin{figure}[htp]
\centering
\includegraphics{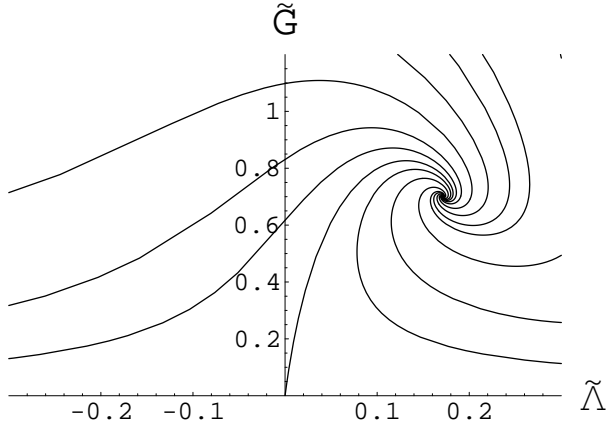}
\caption{The flow in the $\tilde\Lambda$--$\tilde G$ plane in the Einstein--Hilbert truncation.}
\end{figure}

A closely related line of research has to do with the addition of matter.
We have seen earlier that minimally coupled matter fields by themselves induce a nontrivial
FP in the gravity sector. 
It has also been shown that in the Einstein--Hilbert truncation the presence of minimally
coupled matter modifies the position of the FP and the critical exponents,
but asymptotic safety remains a rather generic property of the theory \cite{perini1}.
But does this property persist when we take into account also matter interactions?
Another aspect of this issue is that in the standard model the
the abelian gauge coupling and the scalar self coupling grow with energy.
This indicates that most likely the standard model cannot be a complete theory in itself.
Could gravity fix also this problem?
According to an old conjecture \cite{fradkin} 
all matter interactions become asymptotically free in the presence of gravity.
If this was the case, then in order to establish the existence of a FP for gravity coupled 
to matter it would be enough to consider minimally coupled matter.

Evidence in favor of Fradkin and Tseytlin's conjecture
comes from calculations in \cite{perini2}.
In that paper we computed the beta functions of
theories of gravity coupled to a real scalar with Lagrangian
\begin{equation}
\sqrt{|g|}\left(-\frac{1}{2}g^{\mu\nu}\partial_\mu\phi\partial_\nu\phi 
- V(\phi^2)+F(\phi^2)R\right)
\end{equation}
where  $V(\phi^2)=\sum_{n=0}^\infty {\lambda}_{(2n)} \phi^{2n}$ and
$F(\phi^2)=\sum_{n=0}^\infty {\xi}_{(2n)} \phi^{2n}$, 
possibly in the presence of additional minimally coupled matter fields.
It was shown in \cite{perini2} that, depending on the number of these matter fields, 
the theory has a so--called  ``Gaussian--Matter'' FP, meaning that only the terms
$\lambda_0$ and $\xi_0$ (which correspond to $g_0$ and $g_1$ in the notation of (\ref{expansion}))
are nonzero, while all $\lambda_i$ and $\xi_i$ with $i\geq 1$ (and therefore all
scalar self--interactions) vanish at the FP.
The critical surface of this FP is finite--dimensional.
This means that there exist renormalizable theories of a scalar coupled to gravity, 
which for large but finite $k$ have polynomial self--interactions and polynomial
nonminimal interactions (the degree of the polynomials being determined by the
number of other minimally coupled matter fields in the theory)
and all these interactions are asymptotically free.
Thus gravity seems indeed to heal the UV behaviour of the scalar potential.
Yukawa interactions have been discussed in \cite{zanusso1}.

\subsection{Higher derivative truncations}

The first application of the ERGE beyond the Einstein-Hilbert truncation 
was in \cite{lauscherR2}, where the addition of a term $R^2$ was considered.
Subsequently there have been two significant enlargements of the truncation of the ERGE: 
on one hand the ``exact'' treatment of all four derivative terms,
on the other of polynomials in the scalar curvature up to eighth order.

In order to go beyond the one loop approximation results cited earlier, 
\cite{bms} have reorganized the inverse propagator in terms of Lichnerowicz Laplacians.
Then, they find the following fixed point:
$$
\frac{1}{2\xi_*}+\frac{1}{6\rho_*}=0.00754\ ;\quad
\frac{1}{2\lambda_*}+\frac{1}{\rho_*}=-0.0050\ ;\quad
\tilde\Lambda_*= 0.219\ ;\quad
\tilde G_*= 1.96
$$
with critical exponents 2.51, 1.69, 8.40 and -2.11.
This result has two important implications.
The first is that the couplings $\lambda$ and $\xi$ may not be asymptotically free,
as in perturbation theory, but rather have a finite limit.
The second is that unlike in lower truncations, here not all critical exponents
are positive, and the critical surface, in this truncation, happens to
be three dimensional.

Beyond four derivatives, one can treat Lagrangians that are polynomial in the scalar curvature,
of the form
\begin{equation}
\label{fofr}
\Gamma_k=\int d^4x\sqrt{g}\sum_{i=0}^n g_i R^i
\end{equation}
What makes the calculation feasible is that all these operators scale
differently and therefore can be distinguished by working on a sphere. 
In a suitable gauge, the beta functions for these theories were calculated
in \cite{cpr,masa} and were found to have nontrivial fixed points which
generalize those that were known from lower truncations.
The results are given in tables I and II.
Table \ref{tab:mytable1} gives the
position of these nontrivial FP and table \ref{tab:mytable2} gives the
critical exponents, for truncations ranging from $n=1$ (the
Einstein--Hilbert truncation) to $n=8$. 

\begin{table}
\begin{center}
\begin{tabular}{|c|l|l|l|r|r|r|r|r|r|r|r|}
\hline
 $n$ & $\tilde\Lambda_*$&$\tilde G_*$ &  \multispan9 \hfil $10^3\times$ \hfil
\vline
\\
\hline
  &  & &$\tilde g_{0*}$ & $\tilde g_{1*}$ & $\tilde g_{2*}$ & $\tilde g_{3*}$
& $\tilde g_{4*}$ & $\tilde g_{5*}$ & $\tilde g_{6*}$& $\tilde g_{7*}$& $\tilde g_{8*}$\\
\hline
1&0.130 &0.988  & 5.23& -20.14& & & & & & &\\
2&0.129 &1.563  & 3.29& -12.73& 1.514& & & & & &\\
3&0.132 &1.015  & 5.18& -19.60& 0.702& -9.68& & & & &\\
4&0.123 &0.966  & 5.06& -20.58& 0.270& -10.97& -8.65& & & &\\
5&0.123 &0.969  & 5.07& -20.53& 0.269& -9.69 & -8.03& -3.35& & &\\
6&0.122 &0.958  & 5.05& -20.76& 0.141& -10.20& -9.57& -3.59& 2.46 & &\\
7&0.120 &0.949  & 5.04& -20.97& 0.034& -9.78 & -10.52 & -6.05& 3.42 & 5.90&\\
8&0.122 &0.959  & 5.06& -20.75& 0.088& -8.58 & -8.93& -6.81 & 1.16 & 6.20 &4.69\\
\hline
\end{tabular}
\end{center}
\caption{Position of the FP for increasing order $n$ of the truncation.
To avoid writing too many decimals, the values of $\tilde g_{i*}$ have
been multiplied by 1000.}
\label{tab:mytable1}
\end{table}

\begin{table}
\begin{center}
\begin{tabular}{|c|c|c|c|c|c|c|c|c|c|}
\hline
 $n$ & $Re\vartheta_1$ & $Im\vartheta_1$ & $\vartheta_2$ & $\vartheta_3$
& $Re\vartheta_4$ & $Im\vartheta_4$ &  $\vartheta_6$ & $\vartheta_7$& $\vartheta_8$\\
\hline
1& 2.382& 2.168& & & & & & & \\
2& 1.376& 2.325& 26.862& & & & & &\\
3& 2.711& 2.275& 2.068& -4.231& & & & &\\
4& 2.864& 2.446& 1.546& -3.911& -5.216& & & & \\
5& 2.527& 2.688& 1.783& -4.359& -3.761 & -4.880 & & &\\
6& 2.414& 2.418& 1.500& -4.106& -4.418 & -5.975 & -8.583 & &\\
7& 2.507& 2.435& 1.239& -3.967& -4.568 & -4.931 & -7.572 & -11.076 &\\
8& 2.407& 2.545& 1.398& -4.167& -3.519 & -5.153 & -7.464 & -10.242 & -12.298\\
\hline
\end{tabular}
\end{center}
\caption{Critical exponents
for increasing order $n$ of the truncation.
The first two critical exponents $\vartheta_0$ and $\vartheta_1$
are a complex conjugate pair.
The critical exponent $\vartheta_4$ is real in the truncation $n=4$ but for $n\geq 5$ it becomes
complex and we have set $\vartheta_5=\vartheta_4^*$.}
\label{tab:mytable2}
\end{table}

Looking at these tables, one can make the following observation.
The first is that a FP exists for all truncations considered.
The second is that the properties of the FP
are remarkably stable under improvement of the truncation. In
particular the projection of the flow in the $\tilde\Lambda$-$\tilde
G$ plane agrees well with the case $n=1$. This confirms the claims
made in \cite{lauscher1} about the robustness of the Einstein--Hilbert
truncation.
The greatest deviations occur in the row $n=2$, and in the columns $g_2$ and $\vartheta_2$.
This may be related to the fact that $g_2$ is classically a marginal variable.

The third observation is that in all truncations only three operators are relevant.
One can conclude that in this class of truncations
the UV critical surface is three--dimensional. Its tangent space at the FP
is spanned by the three eigenvectors corresponding to the eigenvalues with
negative real part. In the parametrization (\ref{fofr}),
it is the three--dimensional subspace in ${\bf R}^9$ defined by the equation:
\begin{eqnarray}
\label{surface}
\tilde g_3&=& 0.0006 + 0.0682\,\tilde g_0 + 0.4635\,\tilde g_1 + 0.8950\,\tilde g_2\nonumber\\
\tilde g_4&=&-0.0092 - 0.8365\,\tilde g_0 - 0.2089\,\tilde g_1 + 1.6208\,\tilde g_2\nonumber\\
\tilde g_5&=&-0.0157 - 1.2349\,\tilde g_0 - 0.7254\,\tilde g_1 + 1.0175\,\tilde g_2\nonumber\\
\tilde g_6&=&-0.0127 - 0.6226\,\tilde g_0 - 0.8240\,\tilde g_1 - 0.6468\,\tilde g_2\nonumber\\
\tilde g_7&=&-0.0008 + 0.8139\,\tilde g_0 - 0.1484\,\tilde g_1 - 2.0181\,\tilde g_2\nonumber\\
\tilde g_8&=& 0.0091 + 1.2543\,\tilde g_0 + 0.5085\,\tilde g_1 - 1.9012\,\tilde g_2
\end{eqnarray}
There is a clear trend for the highest eigenvalue to grow with the power of $R$,
so one is justified in believing that no further relevant operators would be
encountered by extending the truncation.
This, together with the result from \cite{bms} that one operator in the
four-derivative truncation is irrelevant, suggests that the critical surface
of pure gravity is three dimensional.

\section{Conclusions}

In the first lecture I argued that at a fundamental level
gravity must be a theory of connections. This is certainly not a new idea.
Theories of gravity with torsion have been around for a long time,
and are often referred to as ``Einstein-Cartan-Sciama-Kibble'' theory.
Ashtekar's reformulation of General Relativity is also based on 
a dynamical connection \cite{ashtekar}.
The $GL(4)$-invariant formulation of the theory that I have described 
has the virtue of exposing the occurrence of a Higgs phenomenon
which makes the connection massive and thus explains
why it is not dynamical at low energy.
This brings gravity much closer to what we know  about the other interactions:
in this picture the reason why we do not see a dynamical gravitational connection
is the same as the reason why we did not see the weak $SU(2)$ gauge fields
until we could construct a sufficiently large accelerator.

A peculiar feature of this Higgs phenomenon is that the
characteristic order parameter is not a scalar but a one form.
(In certain approaches to gravity based on the Plebanski action, 
it could be a two form, but this is equivalent to the soldering form on shell.)
It is usually the case that a nontrivial VEV for a one form would break Lorentz invariance. 
Assuming that the VEV of the theory corresponds to Minkowski space
we can choose bases so that $\theta^a{}_\mu=\delta^a_\mu$, which indeed
breaks the Lorentz transformations regarded as diffeomorphisms of spacetime:
$\theta^a{}_\mu\to \theta^a{}_\nu\Lambda^\nu{}_\mu$.
However there is another realization of the Lorentz group 
acting as a diffeomorphism followed by the inverse transformation in the internal space:
$\theta^a{}_\mu\to \Lambda^{-1a}{}_b\theta^b{}_\nu\Lambda^\nu{}_\mu$
and this is the familiar global Lorentz group that is not broken.

Historically, the point of view discussed here can be represented by Elie Cartan.
It emphasizes the role of fiber bundles and is therefore naturally close in spirit
to the geometrical treatment of gauge theories.
Einstein on the other hand never saw much use for torsion or nonmetricity, and in a sense he was right:
one can describe gravity as we know it perfectly well without them.
In time, this view prevailed, and textbooks on general relativity usually just assume that
the connection is the Levi Civita connection.
If the connection is allowed to be dynamical, it is more often by allowing torsion than
nonmetricity.
The discussion in section 3 should have made it clear that torsion and nonmetricity 
play very similar roles, and from the geometrical point of view there is little
reason to allow one but not the other.
Also, it should be clear that while Einstein's point of view is perfectly appropriate to describe
gravity at low energy, it obscures important structures that could play a role at the Planck scale.

Ultimately the most important reason for adopting Cartan's attitude may be
that it offers a route towards the unification of gravity with the other interactions,
in the strict sense in which this word is used in particle physics.
I have called such a theory a GraviGUT.
Remarkably, it appears that Einstein had at some point contemplated a similar
geometrical scheme in his quest for a unified theory of gravity and electromagnetism \cite{einstein}.
This attempt was abandoned because it failed to reproduce the structure of particles,
but it is clear that key notions were missing then, 
so our modern perspective is quite different.
There are also similarities to the approach proposed in \cite{weinberg2},
the main difference being that here only the dimensionality of the
internal spaces is increased, not that of spacetime.

There are some steps in the construction of a GraviGUT that do not seem to pose
excessive difficulties.
I have described the fermionic sector of one such theory, based on a
Majorana-Weyl representation of the unifying group $SO(3,11)$ \cite{gravigut}.
Some steps in the construction of the bosonic sector have been described
elsewhere \cite{percacci,graviweak,smolin}.

But there are also several outstanding obstacles to the realization of this program.
An open question, already mentioned in section 4, is 
what drives the theory towards the ``broken symmetry'' phase.
In ordinary GUTs it is the shape of a scalar potential;
here things cannot work exactly the same way, for two reasons:
one is that the order parameter is not a scalar, the other that
with a single metric one cannot write a potential with nontrivial minimum.
I will return to this point below.

The second issue, which is bound to appear in any theory of this type is
the problem of ghosts. Theories of gravity
containing two curvatures, as well as gauge theories with noncompact groups
contain massive particles with negative residues.
This is based on tree level analyses, so really nobody knows whether these
ghosts will propagate or not; unfortunately this looks like a
difficult dynamical problem.

The third and perhaps biggest challenge is that a GraviGUT must be based on a 
quantum field theory in four dimensions,
so it will be necessary to somehow overcome the perturbative
nonrenormalizability of Einstein's theory.
Asymptotic safety may be the answer here.
It is based on the assumption that the RG flow of the gravitational couplings
will have a fixed point with finitely many UV attractive directions.
Could this be the case?
I have listed a number of calculations of the gravitational beta functions:
the $\epsilon$--expansion, the $1/N$ expansion, one loop calculations
with generic curvature--squared terms and ``exact'' calculations with truncated 
actions containing up to eight powers of $R$.
Some of these calculations have been repeated in a variety of ways:
using or not using a spin--mode decomposition, with different cutoff functions,
in different gauges. 
At least qualitatively, the properties of the FP are quite insensitive to these choices.
So, while none of these calculations by itself proves that gravity is asymptotically safe,
the broad agreement between all these results provides by now rather
convincing evidence that this may be the case.

All the work done on asymptotic safety of gravity so far is based on the metric
formalism, with the connection constrained to be the Levi-Civita connection.
Thus it does not seem to be directly relevant to the GraviGUT program.
However, every connection can be split into 
the Levi-Civita connection plus a tensor $\Phi$, which behaves pretty much
like any other matter field.
Therefore the asymptotic safety of gravity with a dynamical connection 
is equivalent to the asymptotic safety of gravity with the Levi-Civita connection,
plus suitable matter fields.
In this sense the work done so far is directly relevant to the issue of GraviGUT.
Eventually it will be interesting to discuss directly the beta functions
in formulations of the theory with a dynamical connection.

If this program works, then it seems that we will have a description of gravity
valid up to arbitrarily high energies, always remaining within the
``broken symmetry'' phase.
The topological phase seems to be out of reach, and the question of
what drives the VEV of the order parameter seems less important in this context.
Actually, one has to remember that the ERGE only holds for a bimetric average effective action
$\Gamma_k(g_{\mu\nu},\bar g_{\mu\nu})$, with separate functional dependences
on the background metric and on the ``classical'' metric.
It has been emphasized recently that the inclusion of genuinely
bimetric terms in the truncation could somewhat change the picture
of the flow, although a FP seems still to exist \cite{manrique2}.
A whole class of bimetric terms without derivatives could be
interpreted as potentials for the classical metric.
Similar effective potentials had been studied in the past
\cite{floperc} and in certain cases it was found that they have a minimum
when the classical metric is equal to the background metric.
%The question about what drives dynamically the order parameter could thus
%find a selfconsistent answer.

It is important to understand that the asymptotic safety program 
is a {\it bottom up} approach to quantum gravity.
One starts from the formulation of gravity as an effective field theory
and calculates its beta functions.
One then follows the RG flow towards increasing energy;
if a FP with the desired properties is found,
and if we assume that the real world is described by a trajectory 
ending at the FP, then the theory makes sense up to arbitrarily
high energy. In this sense it becomes a fundamental theory.
The FP action describes the behaviour of gravity from the Planck scale upwards,
and equations such as (\ref{surface}) could in principle be turned into predictions
about scattering amplitudes or other observables.
Of course I am not claiming here that equations (\ref{surface}) are to be taken
literally as the correct predictions: there is too much that we are neglecting
in the calculations. 
However it seems possible that with more effort the asymptotic safety
program will eventually produce realistic predictions.
We do not have the possibility to do scattering experiments at those energies,
so any tests of this theory will probably come from cosmology \cite{bonanno}.
Otherwise, in order to extract low energy predictions from this theory 
one will have to integrate the RG flow, which is likely to prove a challenging task.
But predictions for Planckian physics could have another use.
In order to make contact with low energy physics,
any other ``top down'' approach to quantum gravity will have to extract
from the basic theory an effective field theory, which should agree with the one 
we are describing. It would be very satisfactory if the predictions from
asymptotic safety could be matched by independent arguments.

A final point regards the origin of mass scales \cite{hierarchy}.
The QCD scale can be seen as the scale at which the color gauge coupling
becomes sufficiently strong to drive the formation of bound states.
The Fermi scale is related, in the standard model,
to the shape of a scalar potential.
In the asymptotic safety scenario the Planck scale appears in a different guise.
It is the threshold that separates two very different regimes:
the low energy regime where $G$ does not run and $\tilde G\approx k^2$,
from the fixed point regime where $\tilde G$ does not run and $G\approx k^{-2}$.
I have discussed the similarities between electroweak chiral perturbation theory
(a gauged NLSM) and gravity, regarded as effective field theories.
Recent work reinforces these similarities \cite{codello2,zanusso2}.
This prompts the question whether the higgsless version of the standard model
could be asymptotically safe. (See \cite{gies} for related work.)
The origin of electroweak symmetry breaking will hopefully be clarified by the LHC in the next few years.
The lessons that this will teach us may turn out to be useful also in our struggle to
understand gravity.

\end{document}